\documentclass[11pt]{article}

\usepackage[utf8]{inputenc}
\usepackage[T1]{fontenc}
\usepackage{mathpazo}           
\usepackage[margin=1in]{geometry}
\usepackage{graphicx}
\usepackage{booktabs}
\usepackage{array}              
\usepackage{enumitem}
\usepackage{xcolor}
\usepackage{hyperref}
\usepackage{amsmath}
\usepackage[numbers]{natbib}    
\usepackage{setspace}
\usepackage{parskip}            
\usepackage{caption}
\usepackage{subcaption}

\hypersetup{
    colorlinks=true,
    linkcolor=blue!60!black,
    citecolor=blue!60!black,
    urlcolor=blue!60!black,
}

\newcommand{\eroi}{eROI}        
\newcommand{\compass}{Compass}  

\begin{document}

\title{\textbf{AI Strategy: How to Choose What AI Product to Implement}}

\author{
    Foster Provost\thanks{Stern School of Business, New York University. Email: \texttt{fprovost@stern.nyu.edu}. ORCID: \href{https://orcid.org/0000-0002-0307-3884}{\mbox{0000-0002-0307-3884}}}
    \and
    Panos Ipeirotis\thanks{Stern School of Business, New York University. Email: \texttt{panos@stern.nyu.edu}. ORCID: \href{https://orcid.org/0000-0002-2966-7402}{\mbox{0000-0002-2966-7402}}}
}

\date{\today}

\maketitle

\begin{abstract}
Firms struggle to choose AI projects that pay off: two projects can look equally promising to smart, motivated stakeholders and yet deserve opposite decisions. At the residential real-estate brokerage \compass{}, one AI product (Likely-to-Sell recommendations) flagged sales outreach opportunities and went on to account for nine figures in annual gross commission revenue. Another championed AI product (a Time-on-Market pricing tool) was rightly shelved. A simple ROI estimate could not distinguish the two. We present expected ROI (\eroi{}), a framework that decomposes each bet into three components and rates them separately: \textbf{Value if Successful}, \textbf{Likelihood of Success}, and \textbf{Investment Required}. Each maps to a question executives can answer before building: How valuable would it be if it worked? How likely is it to work? And what would it cost to implement? Separating the three breaks a common catch-22: teams cannot estimate ROI until they know whether a project will work, yet cannot know whether it will work without building it. Judging Value if Successful on its own dissolves the loop, letting a team argue that a product would be valuable if it worked while it weighs how likely that is. The framework also asks, before ranking anything, whether there are enough good ideas on the table. After ranking, it guides assembling a portfolio of bets rather than funding only the single top-ranked project. We illustrate \eroi{} on \compass{}'s candidate AI products. Precise ROI estimates are hard to make given the inherent uncertainty of AI projects. Coarse business-level ratings of the three components are enough to tell strong bets from weak ones.
\end{abstract}


\section{Introduction}
\label{sec:introduction}

In 2019, the residential real-estate brokerage \compass{} had assembled a strong AI team and needed to decide what AI products to implement. The company identified scores of potentially value-adding applications and had to choose where to invest. Two candidates looked comparably attractive to its business, product, and engineering leaders. One, a system that flagged which of an agent's contacts were most likely to sell their homes (Likely-to-Sell recommendations), was implemented and grew into a lasting revenue contributor. The other, a tool that predicted how long a home would sit on the market at a given asking price (a Time-on-Market pricing tool), was set aside. Smart, motivated people championed both, and the simple ROI estimate companies usually demand could not tell them apart: two bets that looked equally worth making, yet deserved opposite decisions.

The puzzle is not unique to \compass{}, and getting it wrong is expensive. A RAND study found that more than 80\% of AI projects fail, twice the rate of non-AI IT projects, often because teams aim AI at targets that are simply too hard \citep{rand_ai_failure}. Generative AI (GenAI) has not broken the pattern: one widely cited industry analysis estimated that roughly 95\% of enterprise GenAI pilots deliver no measurable business impact \citep{mit_ai_pilots}. That headline figure has been debated on methodological grounds, but it is directionally consistent with reports that companies are scrapping AI initiatives at rising rates \citep{spglobal_ai_failure} and that even firms reporting strong returns expect few of their experiments to reach production soon \citep{deloitte_genai_2025}. Some of this risk is inherent to ``science-based'' products (we don't know in advance whether they will work), but much of it traces to a more avoidable cause: poor choices about what to pursue in the first place.

The \eroi{} framework provides a systematic way to rank candidates by \textit{likely} return on investment, rather than by visibility or fashion or first-come-first-served. Its core move is to refuse to collapse a bet into a single ROI figure and instead decompose it into three components rated separately: \textbf{Value if Successful}, \textbf{Likelihood of Success}, and \textbf{Investment Required}. Each maps to a question executives can answer before committing: \emph{How valuable would the product be if it worked? How likely is it to work? And what would it cost to implement?} Firms have lacked such structure, and recent work has begun to supply it, converging on a theme: organizations need structured ways to evaluate AI investments in business terms. \citet{hbr_ai_strategy_value} argue that companies must get the ``strategy-value-AI'' sequence right, starting with a compelling strategy and identifying a leap in buyer value before deploying AI as a tool. \citet{davenport2025enterprise_genai} presses for a shift from ad hoc experimentation to enterprise-aligned efforts with measurable outcomes. \citet{siegel2024measuring} warns that most AI projects report only technical metrics (precision, recall) rather than business metrics (profit, ROI), which often kills projects that could otherwise succeed. An MIT SMR executive guide compiles practitioner playbooks for extracting business value from generative AI rather than adopting it for its own sake \citep{smr_genai_exec_guide}. This work establishes the goal; what it leaves open is the mechanism for choosing among bets.

Keeping the three components apart resolves a common catch-22: a team cannot estimate ROI until it knows whether the project will work, yet cannot know whether it will work without building it. Judging the components one at a time dissolves the loop. And of the three, one usually decides the outcome. What sinks most AI projects is not whether a working product would be valuable or affordable, but whether it can be built at all: whether an AI accurate enough for the application is even achievable. This is the scientific uncertainty behind the ``science-based'' products of our opening, the risk we cannot resolve in advance without doing the work. Rating that likelihood on its own lets a team weigh the bet before committing to it. Precise numbers are rarely necessary: coarse ratings, for example Low, Medium, High, or Very High on each component, carry enough information to rank projects and build confidence in the choice. Each component can be subdivided when an assessment needs it.

Before ranking anything, the framework asks whether there are enough good ideas on the table at all: a firm that has not enumerated widely may hold no likely winner in its consideration set, a lapse cheap to fix but fatal when skipped. In our experience the opposite problem is more common: work with experienced AI professionals to enumerate widely, and the set of value-adding ideas soon grows too large to fund.

We illustrate the framework on the candidate AI products of \compass{}, the brokerage from our opening example: the two bets above and a third, a generative-AI tool for producing renovation visualizations of listings. \compass{} is especially instructive because we can examine both the choices made in 2019 and what happened in the years since, including one product later credited with nine figures in annual gross commission revenue \citep{compass_earnings_2021q4}. The case is rooted in residential real estate, but the principles are not: any organization that must choose what AI projects to invest in faces the same prioritization problem, and the same framework applies. We actually used the framework to help prioritize \compass{}'s AI investments at the time, so the 2019 cases reflect it in use rather than a post hoc reconstruction. Because we were among the decision-makers and write with hindsight, we offer all three to show how the framework organizes such choices, not as an independent test of its predictive accuracy. The next section presents the \eroi{} framework and relates it to existing methods; we then turn it into an organizational process for building a portfolio, and work all three \compass{} projects through it.

\section{The \eroi{} Framework}
\label{sec:framework}

The \eroi{} framework breaks expected return on investment into three components: the \emph{value} a project would create if it succeeded, the \emph{likelihood} that it does, and the \emph{investment} it requires. Estimating the three separately, rather than collapsing them into a single ROI number, is the framework's central move, and it pays off before any number is computed.

Keeping the components separate defuses the same catch-22 where funding is on the line: teams must claim substantial potential ROI to secure it.  Holding value apart from likelihood lets a team argue that a project would deliver high ROI \emph{if successful}, and assess its probability of success on its own terms.  If there is too much uncertainty for the likelihood of success to be rated as High, the team can propose a variety of small investments (R\&D studies, proofs of concept, pilots) to reduce the uncertainty. Framing the components in business terms rather than technical terms also addresses a failure mode \citet{siegel2024measuring} identifies: teams that judge projects by precision and recall rather than profit and revenue leave executives unable to tell whether an investment is worthwhile.

The separation helps in a second way: many stakeholders do not naturally reason in terms of expectations (returns weighted by probabilities), and the decomposition lets them deliberate clearly about how to estimate each component and how it should bear on the decision.

These three components combine as:

\begin{equation}
\label{eq:eroi}
\textit{\eroi{}} \;\propto\; \frac{\text{Value if Successful} \times \text{Likelihood of Success}}{\text{Investment Required}}
\end{equation}

We write it with a proportionality sign rather than an equals sign to signal that a project can be attractive in more than one way.  A project is worth pursuing when a large payoff is both likely and cheap to chase.  But a quick win may also be attractive: perhaps lower value if successful, yet very likely to succeed and requiring little investment.  So may a moonshot: \textit{very} high value if successful, but perhaps high investment and low-to-moderate likelihood of success.  We will discuss how to deal with lower likelihoods of success later.

\begin{figure}[t]
    \centering
    \includegraphics[width=\textwidth]{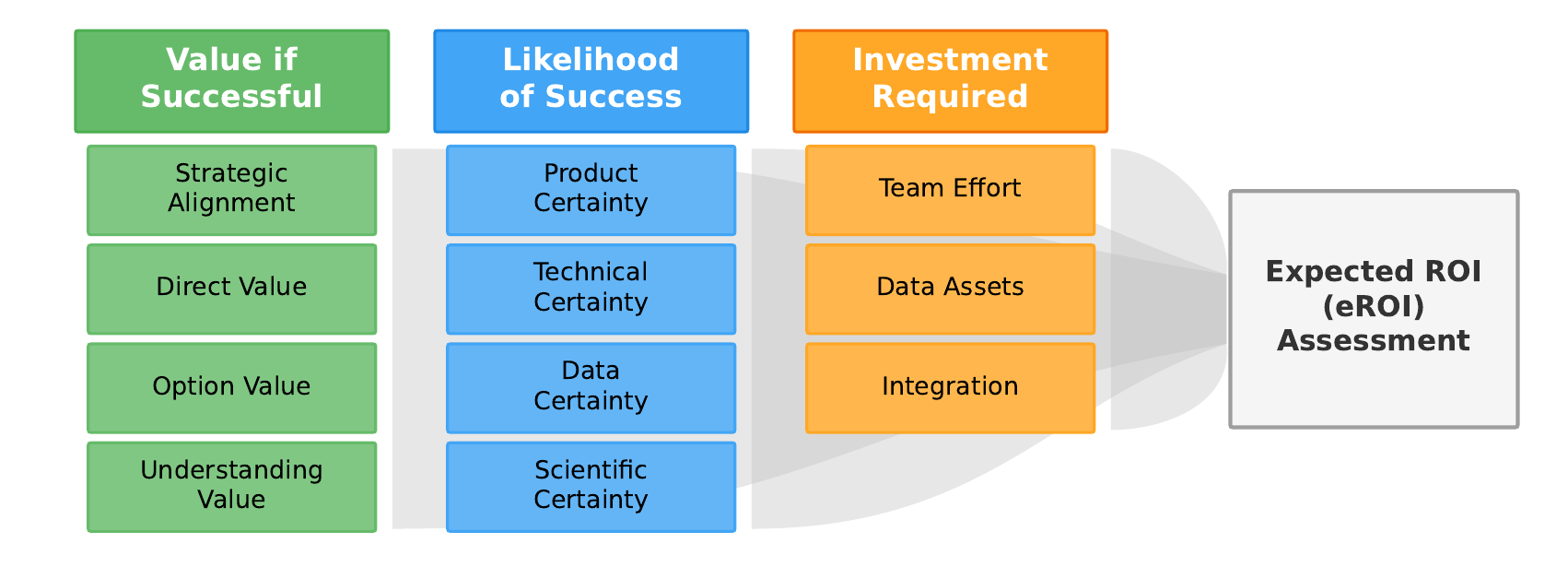}
    \caption{The architecture of the \eroi{} framework. Expected ROI is decomposed into three pillars (Value if Successful, Likelihood of Success, and Investment Required).  Each pillar has several distinct sub-dimensions. By isolating the scientific certainty that an accurate-enough model can be built at all from ultimate business value, organizations can compare diverse AI initiatives on the same terms.}
    \label{fig:eroi-architecture}
\end{figure}

Figure~\ref{fig:eroi-architecture} illustrates the architecture of the framework. Each of the three components, the \emph{pillars} of the framework, breaks down into sub-dimensions, allowing teams to pinpoint where a project is strong or weak rather than relying on a single gut-feeling estimate.

The proportionality sign carries a second message: we usually need not compute the precise quantity in Equation~\ref{eq:eroi}. We have found that ordinal ratings often present an even more compelling case than a single estimated ROI. We rate each component on a coarse scale (Low, Medium, High, or Very High) and judge projects by their \emph{profiles} across the sub-dimensions; the final prioritization is a judgment call by the stakeholders, informed by the full profile rather than a mechanical average of it. For example, we usually need to draw in additional strategic and managerial considerations to decide how to prioritize a quick win versus a more substantial potential blockbuster.

The sections that follow take each pillar in turn, examining the sub-dimensions a team should weigh.

\subsection{Component 1: Value if Successful}
\label{sec:value}

Value has several distinct dimensions.  Focusing too narrowly on direct product value can cause organizations to miss the option value and understanding value a project can create.

\paragraph{Strategic Alignment.}
Before assessing any other dimension of value, we must ask: how will this project help the company achieve its strategic goals? Alignment is the first filter. If a project's strategic fit takes long explanations to justify, or its benefits are too indirect, we should question whether it is the right investment. This is more than good hygiene: firms that capture value from AI draw most of it from core business functions \citep{bcg_value_ai2024}.  Strategic fit and realized value tend to move together.

\paragraph{Direct Product Value.}
This is the business benefit the AI product delivers to customers or the organization.  For revenue-enhancing products this is normally an estimate of the steady-state profit the product or service will produce.  For cost-reduction products we need to consider the net cost reduction, and think end-to-end. By analogy with Amdahl's law \citep{amdahl1967} (which holds that speeding up one stage of a process improves the whole only as much as that stage's share of the total), automating a single step lifts the overall workflow only in proportion to how much of it that step governs. In addition, many AI cost-reduction solutions require business process reengineering, including new roles for employees (for example, verifying the accuracy of AI-generated content).  The cost of the reengineering would be part of the Investment component of the \eroi{}.  But the ongoing cost of the new roles detracts from the net cost reduction.

\paragraph{Option Value.}
Option value comprises the future opportunities a project opens up. AI projects frequently carry such value, because the work done preparing for the project creates assets that can be used for other purposes as well.  It is not unusual for option value to exceed direct product value \citep{provost2013data}. 

Two dimensions deserve attention:

\begin{itemize}[leftmargin=*]
    \item \textbf{Data Asset Value:} AI projects typically create, extend, or refine data assets. Once built, those assets open up options for other products and services that were previously unavailable, offering value far beyond the specific product at hand.
    \item \textbf{AI Platform Value:} When taking a platform approach, every data-science component can support multiple products, so a component's value extends beyond its first application to every other product or service it later enables.
\end{itemize}

\paragraph{Understanding Value.}
Data science can sharpen understanding of the business: testing hypotheses about what drives outcomes and surfacing patterns no one anticipated. Such insight does not always translate immediately into product value, but often yields strategic advantage and new opportunities. Unlike option value, which lives in reusable data and platform \emph{assets}, understanding value lives in \emph{knowledge} that shapes strategy even when it is never shipped.

Assessing the value if successful forces one more choice, and it sets up the next component. What level of success are we arguing for? Do we claim the product will be wildly successful, or would a more measured level of success serve better? The choice matters because it sets the bar for the likelihood of success: a wildly successful product is less likely than a moderately successful one. We have found a sweet spot: the minimum (reasonable) level of success that is nonetheless compelling to the business. The case studies below put this to work.

\subsection{Component 2: Likelihood of Success}
\label{sec:likelihood}

The main reason AI strategy is difficult is that AI projects face far higher uncertainty than the projects decision-makers typically have experience with.

That uncertainty has at least four sources, and we must weigh each. Two conventions govern how the four combine into the overall likelihood-of-success estimate. First, likelihood is \emph{gated} by its weakest necessary condition: each of the four sources is necessary, so a single source of high uncertainty caps the overall likelihood even when the others show relative certainty, and any further weak source only lowers the estimate. Second, we rate each source not by its uncertainty but by its complementary \emph{certainty}, on the same Low-to-Very-High scale, so that, as with value, a higher rating is more favorable.

\paragraph{Product Certainty.} Would the product realize its envisioned value if we built it? Would customers use it and find it valuable?  AI-specific product risks involving trust, safety, and compliance live here, alongside more familiar product uncertainty factors.  A system that raises fairness, privacy, or regulatory concerns, or that users simply do not trust, may fail to win adoption or approval even when it performs as designed, which lowers product certainty. 

\paragraph{Technical Certainty.} Can we build the product cost-effectively with available engineering resources and infrastructure?  Technical uncertainty should be familiar to a firm that regularly implements technical products and services.  AI products bring additional technical uncertainty, especially to firms with little experience in AI algorithms, architectures, and DevOps (``MLOps'').

\paragraph{Data Certainty.} Do we have the data necessary to build the scientific solution? Is the data quality sufficient? Do we clearly understand the target of the AI inference? Are we dealing with proxies rather than direct measures?  A well-known key to success with AI projects is getting the data right.

\paragraph{Scientific Certainty.} How likely are we to achieve a sufficient level of accuracy with the AI solution for the project to be successful?  All AI systems make mistakes.  Will the cost of the errors render the product useless? This is the ``can we even do this in the first place, and how well?'' dimension that distinguishes AI from many other product investments. Do we have compelling hints from initial exploratory data analysis or pilot studies that a sufficiently accurate model is feasible? Data certainty asks whether we \emph{have} the right inputs; scientific certainty asks whether, even with them, the target can be estimated with a low enough error rate.

Two distinct failure modes hide under this one subcomponent. The first is ordinary \emph{predictive} difficulty: even with the right inputs, the pattern may be too weak to estimate to a useful accuracy. The second is subtler, arising when we are deciding to take actions that will have \emph{causal} effects (rather than reacting to a pattern we have detected). Evaluating a causal-effect prediction requires knowing what would have happened otherwise, the counterfactual, which is never observed for any individual unit \citep{holland1986, imbens_rubin2015}. An \emph{average} effect across a group can sometimes still be recovered when the right design is available (an experiment, or a natural experiment that supplies a comparable control group); absent such a design, observational data alone cannot reveal it. Either way this adds a layer of complexity in estimation and decision making \citep{fernandez2022causal}, requiring more complex data collection and elevated statistical expertise; Section~\ref{sec:tom} returns to it.

With generative AI systems, the predictive difficulty is hidden.  The generated responses are produced by a predictive, machine-learned model.  The systems make mistakes just like any AI system.  However, we tend not to think of the generated result as the model's \textit{prediction} of what would be the best response to the prompt.  Because the complex output of the GenAI system results from many inference runs of the underlying model,\footnote{For example, the underlying LLM for text generation or de-noising model used in an image diffusion process} the mistakes are much more difficult to identify automatically.  Nonetheless, scientific uncertainty matters just as much.  For many tasks, we do not know in advance whether a GenAI system output will be accurate enough for a satisfactory product, and how much the errors will cost us (which is needed for an ultimate ROI estimation).  

Scientific uncertainty isn't unique to AI.  For example, pharmaceutical firms assess candidate drugs by their odds of clearing scientific and regulatory hurdles \citep{dimasi2001, dimasi2003}.  Nevertheless, scientific uncertainty must be assessed explicitly: it is the extra ``can we even do this?'' layer on top of the more typical product and engineering risk.

\subsection{Component 3: Investment Required}
\label{sec:investment}

Investment is the pillar we want low rather than high: it sits in the denominator of Equation~\ref{eq:eroi}, so a smaller commitment raises \eroi{}. Teams that budget only for building the model underestimate what a project truly costs. The cost goes well beyond the model, and three kinds of investment matter.

\paragraph{Data Science Team Effort.} How large a job is this for the available technical team? Is it a straightforward application of established techniques, or does it require research and experimentation?

\paragraph{Data Asset Investment.} Do we need to invest in acquiring additional data assets? Can we start with existing data for version 1.0, or are new data sources essential for success?

\paragraph{Integration and Productization Effort.} How much effort is required to integrate the AI capability into existing products and workflows? This includes product and UX integration, ML engineering for pipelines and infrastructure, business process reengineering, guardrailing and monitoring systems, and potentially more humans in the loop. Applications in sensitive or regulated domains carry an extra share of this work: legal and compliance review, fairness and privacy safeguards, and disclosure or consent mechanisms. This ethical and regulatory exposure can raise the investment required well above what an R\&D effort, prototype, or proof of concept would suggest.

\subsection{Putting Them Together}
\label{sec:combining}
With the three pillars in hand, two rules govern how the ratings come together. The first concerns how sub-dimensions combine within a pillar, and Value and Likelihood combine in opposite ways. For Value, the overall rating is a \emph{holistic} judgment informed by the sub-dimension ratings rather than a mechanical sum or average of them: the sub-ratings make the holistic call easier, not automatic. (Some practitioners prefer instead to sum the sub-scores.) Likelihood works the other way. As Section~\ref{sec:likelihood} explains, it is \emph{gated}: a single Low on any of the four certainties, product, technical, data, or scientific, caps overall likelihood even when the others are High, and additional Lows only pull it lower. Investment Required combines as Value does: a holistic judgment over its sub-dimensions rather than a sum or average, so a single High does not by itself raise the overall investment rating when the other dimensions are lower.

The second rule concerns one fact bearing on two pillars: it may inform both without being double-counted, because each pillar scores a distinct conditional question.  For example, a cheap incumbent solution both shrinks the \emph{incremental} payoff (Value) and makes adoption less certain (Likelihood).

\medskip
\noindent Table~\ref{tab:rubric} gives coarse anchors for the three pillars. Shared anchors help raters converge on similar scores, so that disagreement surfaces hidden assumptions rather than mere optimism.

\begin{table}[t]
\centering
\caption{Coarse rating anchors for the three \eroi{} pillars. As noted at Equation~\ref{eq:eroi}, the ratings are ordinal, and a pillar's overall rating reflects the profile of its sub-dimensions (Figure~\ref{fig:eroi-architecture}). For Value and Likelihood a higher rating is more favorable; for Investment Required a \emph{lower} rating is more favorable, since a smaller commitment raises \eroi{}.}
\label{tab:rubric}
{\small
\renewcommand{\arraystretch}{1.25}
\begin{tabular}{@{}>{\raggedright\arraybackslash}p{0.12\textwidth} >{\raggedright\arraybackslash}p{0.185\textwidth} >{\raggedright\arraybackslash}p{0.185\textwidth} >{\raggedright\arraybackslash}p{0.185\textwidth} >{\raggedright\arraybackslash}p{0.185\textwidth}@{}}
\toprule
 & \textbf{Low} & \textbf{Medium} & \textbf{High} & \textbf{Very High} \\
\midrule
\textbf{Value if Successful}\newline\textit{\small (higher is better)} & Marginal gain over the best existing alternative; narrow use; little reuse. & Clear benefit to one product area; some strategic fit; limited option value. & Substantial cross-product benefit; strong strategic fit; creates reusable assets. & Strategy-defining; large direct value plus major option and understanding value. \\
\addlinespace
\textbf{Likelihood of Success}\newline\textit{\small (higher is better)} & Low certainty on any of the four sources: a good-enough outcome is speculative. & At least one of the four sources still unproven; a pilot is needed. & High certainty on all four sources: well-trodden approach, few open risks. & Very high certainty across all four: proven on closely comparable problems. \\
\addlinespace
\textbf{Investment Required}\newline\textit{\small (lower is better)} & Established techniques, existing data, light integration. & Substantial but well-understood build; some new data or integration. & Significant research or productization; new data assets; heavy integration. & Open-ended, research-grade build; major new data assets and infrastructure. \\
\bottomrule
\end{tabular}
}
\end{table}

\subsection{How \eroi{} Relates to Existing Methods}
\label{sec:relation}

\eroi{} adds a decomposition tailored to the distinctive risk of AI, one that does two things a general-purpose score leaves blended. First, it lets stakeholders discuss cleanly what the product would be worth if it succeeded. Second, it isolates \emph{scientific} certainty, distinct from ordinary engineering risk: whether an accurate-enough AI system can be built, that is, whether the target is \emph{learnable at all} to a useful accuracy. Lightweight feature-ranking schemes fold that question into a single ``confidence'' rating. The probability-weighted portfolio score folds it into one ``probability of technical success.'' Neither surfaces it. A corollary of the same separation: option value, the data assets and reusable components a project creates, becomes a first-class dimension rather than a footnote to expected value.

\eroi{} builds that decomposition from familiar ideas. Product teams rank features with lightweight schemes such as RICE \citep{rice_mcbride2018} or ICE \citep{ice_ellis2017}; finance evaluates staged commitments through real options \citep{realoptions_dixit1994, realoptions_trigeorgis1996}; and new-product development screens projects through stage-gate reviews \citep{stagegate_cooper1990}, the staged go/no-go discipline \eroi{} adopts for AI bets. Closest in spirit is the long tradition of R\&D and new-product portfolio management (of which pharmaceutical pipeline valuation is one instance), which ranks uncertain projects by value weighted by the probability of technical success, together with the broader machinery of decision analysis \citep{cooper_portfolio2001, howard1988}. \eroi{} specializes that decision-analytic machinery, supplying the structured value, likelihood, and investment dimensions an analyst would otherwise have to assemble case by case.

A single blended ``confidence'' or ``probability of technical success'' score buries what \eroi{} surfaces. We show as much below on two \compass{} cases: the Time-on-Market tool, which turns on scientific certainty, and the renovation tool, which turns on option value. Both are worked through in full later.

\section{From Framework to Strategy: Guiding Principles}
\label{sec:strategy}

Evaluating individual projects is one thing; applying \eroi{} across an organization requires a deliberate process integrating business, product, engineering, and scientific perspectives. That process has two steps: enumerate candidates widely, then prioritize them into a portfolio.

\subsection{Step 1: Enumerate a Large List of Potential AI Projects}
\label{sec:enumerate}

This stage pairs knowledge of the business's challenges and opportunities with knowledge of what AI can do. Because a company usually has many possible ways to create value with AI, talk to as many stakeholders as possible: not just founders and executives, but product managers, engineering managers, and especially customers. Consistent with \citet{hbr_ai_strategy_value}, enumeration should be driven by business needs and buyer value, not by what the technology can do.

Having too few candidates is a common failure mode. A short list very likely omits great opportunities, sometimes the best ones. Building the list takes effort but no special expertise: managers and customers have usually thought hard about how AI might improve their own area of responsibility. If the firm has someone with real experience across AI successes and failures, empower them as an ``internal consultant'' to help brainstorm the enumeration. A team that does not enumerate widely will not know whether its best opportunities are even on the table, and even a carefully assembled portfolio of AI investments will then underperform.

Do resist the temptation to limit this exercise to generative AI. Although McKinsey estimates that generative AI will add \$2.6--4.4 trillion in annual value globally, this would increase the total impact of all AI by only 15--40\% \citep{mckinsey_genai}. The larger, remaining share is where most of AI's economic value still sits: analytical AI techniques such as predictive machine learning, causal decision making at scale, optimization, and classification. Organizations that focus exclusively on generative AI may overlook their highest-\eroi{} opportunities.

\subsection{Step 2: Prioritize, Then Build a Portfolio}
\label{sec:prioritize}

Prioritizing the enumerated list into a funded portfolio does not mean funding the single highest-ranked project. These are risky investments, so build a portfolio: a deliberate mix of bets beats staking everything on the top-ranked project, because funding then rarely hinges on a single razor's-edge ranking.

Building that portfolio starts with rating each candidate across the three pillars. The aim is rating, not precise calculation: rough ratings suffice, because the \eroi{} analysis is an input to a selection procedure in which decision-makers layer on criteria it cannot capture. A CEO may weigh the relative strategic alignment of two projects in ways the analytic team cannot quantify, and funding and shared resources add constraints of their own: two projects with similar, stellar \eroi{} may need the same scarce AI team, so the firm funds one alongside other bets whose \eroi{}s are attractive if not as stellar. Ratings inform the choice; they do not make it. 

Comparisons are easiest when one profile dominates another (better or equal on every pillar), which coarse ratings make common. When profiles instead conflict (one project higher on value, another lower on investment), \eroi{} generally will not settle the trade: stakeholders must judge what \eroi{} profile is desired. What the framework supplies is structure for that judgment, keeping the opposing dimensions explicit rather than burying them in one score.

A good portfolio spans several archetypes:

\begin{itemize}[leftmargin=*]
    \item \emph{Potential big wins} (Very High value, High likelihood, Medium investment), the profile of \compass{}'s Likely-to-Sell recommendations, which we work through in the case study below.
    \item \emph{Quick wins} (High value, High likelihood, Low investment).
    \item \emph{Strategic bets} (Very High value, Medium likelihood, High investment).
    \item \emph{Research investments} to reduce uncertainty for potential high-value projects (High value, High Uncertainty). We say ``Uncertainty'' rather than ``Low likelihood'' to stress the difference between not knowing whether we can do it and being fairly certain that we cannot. \compass{}'s Time-on-Market tool is the case below.
\end{itemize}

For a \emph{strategic bet}, ``High investment'' describes the eventual cost profile, not a lump sum committed up front: when risks remain (Medium likelihood), the first move is a staged commitment, a pilot or research phase that buys down uncertainty before the firm decides whether to invest further. A \emph{research investment} can look similar, but for a causal-prediction problem such as the Time-on-Market tool it is not an ordinary pilot: the research is a hunt for the identifying variation the prediction needs, and absent that variation the project fails the gate rather than becoming a funded research bet. We work through both \compass{} cases below.

This portfolio discipline echoes recent findings on extracting value from generative AI: firms seeing early returns pursue ``small t'' transformations that build capability over time \citep{smr_genai_smallt} and apply a cost test, using an LLM only where doing so costs far less than business as usual \citep{smr_llm_guide}.

\paragraph{Resisting score inflation.} Because the ratings inform funding, project champions have an incentive to inflate them, a pressure the original 2019 decisions faced (our retrospective retelling does not) and one any prospective decision maker must manage. A few practices help. Have the process run by an independent innovation leader, with a stake in success but not in any particular product; and have a cross-functional panel (business, product, engineering, and data science) rate each pillar independently, then reconcile, so no single advocate sets a project's ratings. Require a careful, evidence-backed case to unlock the top ratings: a \emph{High} or \emph{Very High} on Likelihood should rest on a pilot result, a comparable deployed system, or concrete signals from exploratory analysis, not optimism alone, and a top Value rating should rest on a compelling business case. Treat divergent scores not as noise to be averaged away but as a prompt to surface the assumptions behind them. The goal is not numerical precision but a shared, defensible view that survives cross-examination.

Revisit the portfolio as priorities shift, so promising ideas do not drift to the back of the queue and strategically stale ones do not stay highly ranked. Selection is not the whole job: returns still depend on complementary investments, the right data assets, an experienced AI science and engineering team, and AI product management that can ship and operate AI products successfully. Enumerate widely, prioritize into a portfolio, and defend the ratings that drive it: the case study that follows shows the framework doing exactly that at \compass{}.

\section{Case Study: Applying \eroi{} at \compass{}}
\label{sec:case-study}

We illustrate the framework with \compass{} Inc., a technology-focused residential real-estate brokerage and, by the standards of its industry, an unusually large one: it reported record revenue of \$7.0 billion for 2025 and in January 2026 combined with Anywhere Real Estate \citep{compass_fy2025}.\footnote{\compass{} Inc.\ went public in April 2021; its 2025 revenue came with positive adjusted EBITDA but a small GAAP net loss. The Anywhere combination was an all-stock deal.} In the United States, brokerages represent home buyers and sellers through licensed agents. That scale makes \compass{} a useful setting for the framework: a company with the resources to fund many AI bets and the need to choose among them.

By 2019, \compass{} was building a platform to support and integrate all of an agent's activities, bringing a modern technology perspective to a traditionally technology-light industry, as it pursued its goal of becoming the country's largest independent brokerage.\footnote{Two exceptions to the industry's ``technology-light'' character: the Multiple Listing Services (MLSs), where agents post and find listings, and Zillow, which offers a similar consumer platform.}

\paragraph{\compass{}'s Strategy.}
The company's strategy centered on helping \compass{} agents \emph{get more business} (acquiring more seller or buyer clients) or \emph{get more time} (to spend with their families or other priorities). This clear strategic focus became a filter for evaluating AI investments.

\subsection*{AI Areas of Opportunity}
\label{sec:opportunities}

Extensive discussions with stakeholders across the organization identified scores of potential AI products and services, spanning the company's major product areas:

\begin{itemize}[leftmargin=*]
    \item \textbf{Compass.com} (consumer and agent website): ranking search results, recommending homes, home price estimation, image-based search, auto-generation of listing descriptions, data quality assurance, and more.
    \item \textbf{Marketing Center} (creating marketing assets for agents): improving targeting of mail campaigns, email and social media optimization, learning what creative approaches work best in different situations.
    \item \textbf{CRM System} (managing agent contacts): building relationship graphs, helping agents farm new leads, identifying which contacts to reach out to, creating maps of likely-to-sell properties.
    \item \textbf{Sales Support} (helping agents win business): providing sellers with what-if pricing tools, improving listing presentations, providing market intelligence.
\end{itemize}

The challenge was not finding uses for AI \citep{mims2020ai}: it was choosing from among many opportunities that seemed promising on the surface, indeed far more than the company could pursue simultaneously.

\begin{figure}[t]
    \centering
    \includegraphics[width=0.8\textwidth]{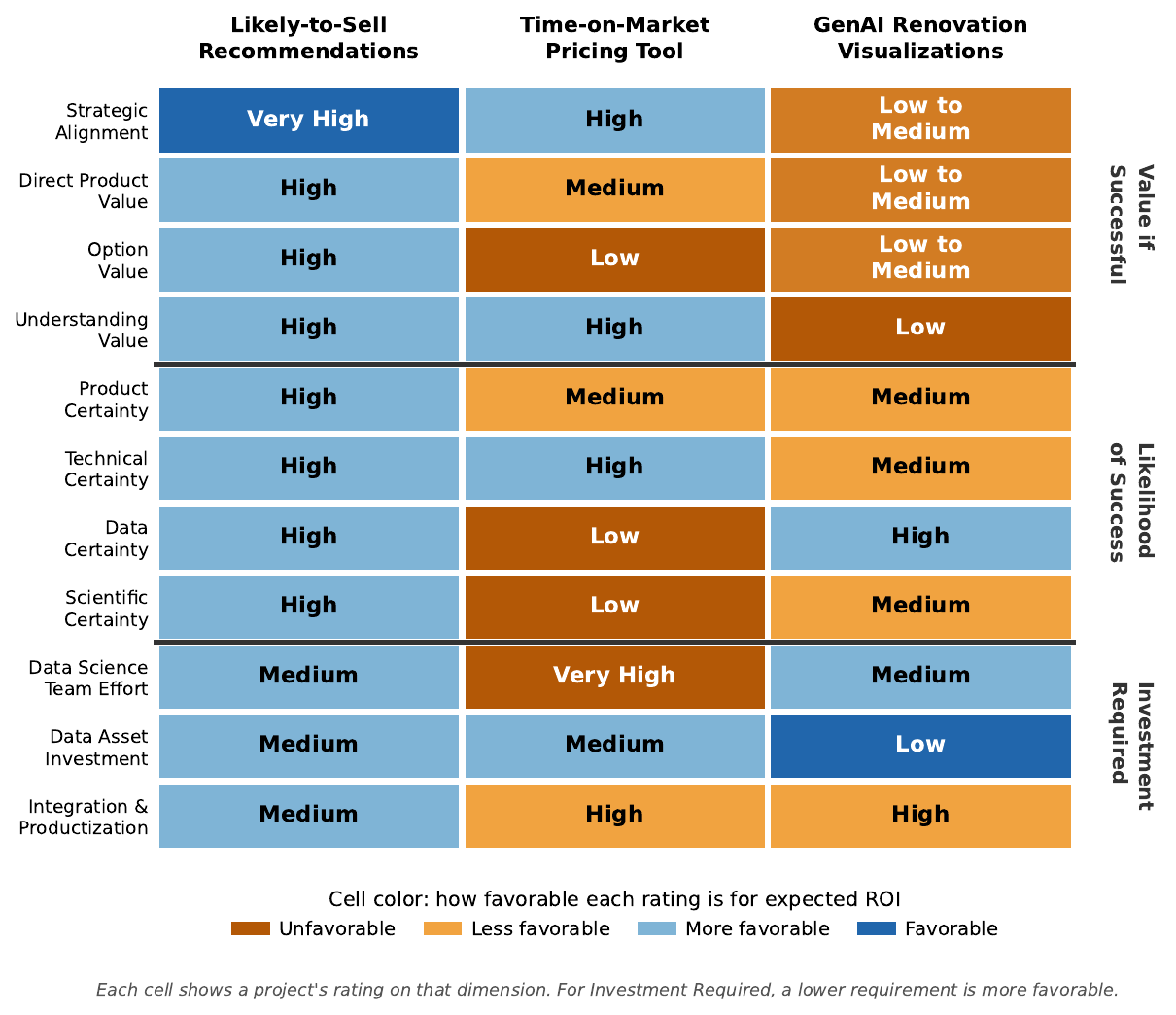}
    \caption{The \eroi{} framework applied to three AI projects at \compass{}. Each cell gives a project's rating on one sub-dimension (Low, Medium, High, or Very High); the cell color shows how \emph{favorable} that rating is for expected ROI, from orange (unfavorable) to blue (favorable), so that blue always indicates higher \eroi{} regardless of the pillar. Because a smaller commitment is better, a low Investment Required is shown blue. Each cell is also labeled with its rating, so the figure does not rely on color alone, and the blue--orange scale is colorblind-safe.
    }
    \label{fig:eroi-visualization}
\end{figure}

To illustrate how the framework guides decisions, we next work through three real \compass{} cases (not hypotheticals): a funded project that succeeded (Likely-to-Sell recommendations), an attractive project that turned out not to be a good choice to prioritize (the Time-on-Market pricing tool), and a generative-AI product (renovation visualizations). Figure~\ref{fig:eroi-visualization} previews how each of the three products scores across every sub-dimension; the following subsections explain the ratings.

\subsection{Example 1: Likely-to-Sell Recommendations for CRM Contacts}
\label{sec:lts}

Residential real-estate agents are in the business of helping consumers buy and sell homes. Much of their business comes from their networks. However, a typical agent has hundreds or thousands of contacts and rarely knows whom to reach out to at any particular time. Thus many agents simply miss getting the business of a contact who puts their home on the market. Likely-to-Sell (LTS) recommendations point agents to the contacts most likely to sell in the near future and help them reach out \citep{compass_lts_product}.

LTS was one of the top-ranked AI projects in mid-2019, as the company planned its 2020 work; its \eroi{} profile shows why.

\subsubsection{Value if Successful}

\paragraph{Strategic Alignment: Very High.} The primary goal of LTS recommendations is to increase every agent's business: identifying relevant opportunities, nudging agents toward them, and facilitating outreach. This is directly aligned with \compass{}'s strategy of helping agents get more business.

\paragraph{Direct Product Value: High.} As the framework advises, we anchor on a reasonable, not overblown, level of success: for LTS, that it adds just one sale per agent per year. As the uncertainties below show, this is not a trivial bar to clear, but it is far from unreasonable. With the average \compass{} listing around \$1 million and commission rates averaging approximately 2.5\%, each added listing brings in about \$25,000 in additional commission revenue. Across \compass{}'s at-the-time approximately 20,000 agents,\footnote{\compass{}'s approximate total agent count (including agents who are team members) during the case-study period, larger than the ``principal agents'' figure in its financial reporting.} that one extra sale apiece implies around \$20 billion in additional annual real-estate transaction volume and roughly half a billion dollars of additional commissions (such commissions are \compass{}'s top-line revenue, split between the agent and the brokerage).

\paragraph{Option Value: High.} The models built for the LTS product help far beyond CRM recommendations. Many likely-to-sell homes aren't owned by \compass{} contacts, suggesting further lead-generation/sales-support products. And the infrastructure built can support other agent recommendations: contacts with rising activity, consumers likely to buy or likely to refer business.

\paragraph{Understanding Value: High.} Understanding the sales dynamics of properties increases the company's ability to create complementary products serving agents, sellers, and buyers.

\paragraph{Overall Value Assessment: \textbf{Very High}.} Strategic Alignment is Very High and the other three dimensions are High; the direct fit with the core strategy and the breadth of follow-on products lift the profile to Very High, a judgment over the profile, not an average.

\subsubsection{Likelihood of Success}

\paragraph{Product Certainty: High.} Agents repeatedly named not knowing which contacts to call as a challenge. LTS focuses agents on their own homeowner contacts, rather than cold-call leads. This should result in a relatively high likelihood that the agent will be willing to initiate a conversation with the homeowner, as well as a relatively high likelihood that the lead will be willing to talk to the agent. A large portion of the agents use the CRM regularly, so product adoption is relatively light (clicking on an additional tab in the CRM system). And even if the homeowner is not ready to sell their home immediately, the conversation will bring the agent (back) into the consciousness of the homeowner, and thus into the consideration set if/when they are interested in talking about selling.\footnote{Note that the vast majority of homesellers talk to only one agent when choosing who to represent them~\citep{NAR_profile}.}

\paragraph{Technical Certainty: High.} The team could build the tool with resources on hand; the existing, well-used, industry-leading \compass{} CRM system was a very suitable place to house it, and the CRM technical team was excited about adding this functionality.

\paragraph{Data Certainty: High.} Detailed lists of all prior property sales were available as training instances. The majority of agents' contacts could be mapped to the home(s) they owned, so a property-level prediction could be tied to a specific contact to call. Although the existing data did not include direct signals for why people sell (life events, family changes), useful proxies were available: how long owners had been in the home, house size, school district, etc., and how those factors had recently correlated with selling. Demand signals were available too: which property types were sought after, neighborhood price and sales trends, and sale frequency for each property.\footnote{For more on feature engineering for LTS, see https://medium.com/compass-true-north/framing-feature-engineering-for-machine-learning-a-generative-model-of-home-likelihood-to-sell-f96f21d6d6d0.}

\paragraph{Scientific Certainty: High.} The modeling and inference problems were straightforward for an experienced AI team \citep{compass_lts_ml}.

\paragraph{Overall Likelihood Assessment: \textbf{High}.}

\subsubsection{Investment Required}

\paragraph{Data Science Team Effort: Medium.} The modeling and evaluation work was a substantial but well-understood job for the team.

\paragraph{Data Asset Investment: Medium.} The team could start with existing data for version 1.0 and invest in additional data assets later, to possibly improve LTS prediction accuracy.  

\paragraph{Integration and Productization Effort: Medium.} The CRM product and engineering team was willing to devote resources to product integration, which aligned with their interests.

\paragraph{Overall Investment Assessment: \textbf{Medium}.}

\subsubsection{The Upshot}

We can now see why this was a top priority: Very High value across multiple dimensions, High likelihood of success, and only Medium investment. By construction, that results in a very high \eroi{}.

\textbf{The Result:} \compass{} invested in building, integrating, and marketing LTS recommendations, and the product became a lasting part of the platform. In its 2021 Q4 earnings call, the company reported a nine-figure sum in annual gross commission revenue from listings the tool had flagged before they came to market \citep{compass_earnings_2021q4}.

This was \emph{incremental} revenue (attributable to the recommendations rather than to listings agents would have won anyway). Measuring that increment after the fact is exactly what the Time-on-Market tool could not do, as the next example shows; here it was possible because the gradual rollout of the recommendations created a natural experiment. Because each agent received only a couple of recommendations per day, many to-be-recommended contacts had not yet been surfaced when they listed their homes. Among contacts the model scored as comparably likely to sell, we have a treatment group (those listed homes where a recommendation was given) and a control group (those not-yet-recommended homes). Comparing these groups, we see not only the market share \compass{} won after recommending, but also the market share \compass{} won on comparable homes \emph{without} a recommendation \citep{imbens_rubin2015}. Measuring the share \compass{} won on \emph{surfaced} recommendations versus that baseline isolates the additional business the recommendations brought in.

As with all causal conclusions, the estimate rests on assumptions. In this case, the principal assumption is that (setting aside the impact of the recommendation itself) at similar model scores, listing just before rather than just after being surfaced is unrelated to whether a seller chooses a \compass{} agent. Even so, Likely-to-Sell is a case where a product's causal \emph{impact}, not merely its predictive accuracy, could be measured. This available, observable counterfactual contrasts with the case for the Time-on-Market tool, as we discuss in Section~\ref{sec:tom}. The LTS product has continued to evolve and expand, building on the AI platform components and data assets created in the initial investment.

\subsection{Example 2: Time-on-Market Pricing Tool}
\label{sec:tom}

Contrast LTS with another application that drew strong stakeholder interest in 2019: a ``what-if'' tool estimating how long a property would sit on the market at a given asking price, helping agents weigh a seller's wish to sell quickly against the higher price a longer wait might bring.

\subsubsection{Value if Successful}

\paragraph{Strategic Alignment: High.} Such a tool would help agents price their clients' listings and focus their effort.

\paragraph{Direct Product Value: Medium.} In the best case, the product could eliminate listings that expire unsold. The benefit of selling faster is harder to quantify, as the value of time varies by agent and client. We size this generously, on its best case, and the choice is deliberate: Likely-to-Sell was credited only a modest one extra sale per agent per year, whereas we grant the Time-on-Market tool its optimistic ceiling. The asymmetry favors the rejected project, so any ranking that survives it is on firm ground. Of the roughly 30\% of listings that do not convert, unrealistic pricing accounts for perhaps three-quarters (about 22.5\% of all listings), so a tool that reliably steered sellers off overpricing would address the single largest cause. Converting even a fifth of those would add roughly one extra sold listing for every five or six agents per year. At about \$25,000 in commissions per listing, that is on the order of \$90 million across \compass{}'s 20,000 agents. Even credited its best case, the tool lands far below the modestly credited Likely-to-Sell (roughly half a billion dollars), though above the renovation tool discussed next; the ranking therefore holds \emph{a fortiori}.

\paragraph{Option Value: Low.} The team would need to brainstorm other applications, so the value here was uncertain, though building the team's capabilities in causal prediction (defined under Scientific Certainty below) had some option value.

\paragraph{Understanding Value: High.} Understanding the price--time relationship would likely inform other strategies and tactics.

\paragraph{Overall Value Assessment: \textbf{Medium to High}.}

\subsubsection{Likelihood of Success}

\paragraph{Product Certainty: Medium.} Stakeholders were enthusiastic, but it was unclear whether agents and their clients would trust and act on a model's ``what-if'' pricing advice.

\paragraph{Technical Certainty: High.} Serving such a tool in the company's tech platform was well within the team's engineering capabilities.

\paragraph{Data Certainty: Low.} Rich observational data on past listings existed, but the kind of data needed to isolate what would happen with different prices, i.e., the \emph{causal} effect of price, did not.  (There was little-to-no data on cases where the asking price varied for reasons unrelated to the home itself, and since each home is unique it is difficult to match different-priced homes \textit{ceteris paribus}).

\paragraph{Scientific Certainty: Low.} This is a \emph{causal prediction} task: the AI estimates the effect of an intervention (the asking price). The counterfactual ground truth (what would have happened at a different price) is never observed. The obstacle was structural. Causal effects can be identified with experiments, instruments (factors that shift price on their own, unrelated to the home), or quasi-experimental variation \citep{angrist_pischke2009}, but the team had none of these: no way to randomize asking prices when homes are listed, no instrument shifting price independently of the home. What it had was observational data and pervasive unobserved confounding (hidden factors moving both price and time to sell), which made cause and effect nearly impossible to disentangle here.

It did seem possible to predict time on market for a past pricing decision, but turning it into a ``what-if'' tool would require strong, risky assumptions about the causal structure.

As discussed earlier, for likelihood, having even one ``Low'' source gates the overall assessment. For the Time-on-Market tool, we have two.

\paragraph{Overall Likelihood Assessment: \textbf{Low}.}

\subsubsection{Investment Required}

\paragraph{Data Science Team Effort: Very High.} This was a research-grade problem within AI and data science: the effort would be substantial, with real uncertainty about whether it would yield a usable product at all.

\paragraph{Data Asset Investment: Medium.} Beyond the observational data already on hand, building toward the causal evidence the method needs would take real effort.

\paragraph{Integration and Productization Effort: High.} Turning a research-grade model into a what-if pricing tool that agents and clients could trust would demand careful product and UX work.

\paragraph{Overall Investment Assessment: \textbf{High}.} The research-grade Team Effort (Very High) and heavy Integration (High) set the level; the Medium data-asset investment does not pull the overall down to Medium, since this is a judgment over the profile, not an average.

\subsubsection{The Upshot}

High scientific uncertainty and high investment together led the \eroi{} to rate this project well below other AI products.

Yet the strong strategic alignment suggested a different kind of investment: \emph{research} to clarify the value-adding product idea and, if one existed, reduce the scientific uncertainty through exploration and pilots. The framework thus guides not just yes/no decisions but when a smaller research investment makes sense first: its signature is high value if successful paired with a likelihood capped by scientific uncertainty that targeted research might reduce. For a causal prediction like this one, though, that research means a hunt for identifying variation (a natural experiment or other source of exogenous price movement), not just an ordinary product pilot, since observational data alone cannot reveal the counterfactual. 

The nature of the causal estimation differs fundamentally between the two tools. Likely-to-Sell does not make causal predictions (calling a contact does not change the model's estimate of their likelihood to sell), so its only causal question is the \emph{post-hoc} evaluation of whether the product raised revenue, and there a usable counterfactual was close at hand (Section~\ref{sec:lts}): had the natural experiment's assumptions seemed too strong, a simple A/B test would have sufficed. The Time-on-Market tool instead asks the AI to make a causal \emph{prediction} (list at this price and time on market will be such-and-such), and even assembling the training data would require running price experiments on real homes with real sellers.

Two lessons follow. First, whether or not the AI product itself makes causal predictions, it pays to have sophisticated causal-estimation expertise on the team; assuming that A/B tests are the only route to causal estimates is a real handicap when such tests are costly or impossible (a business-to-business firm, for instance, can rarely withhold a major new AI offering from a random subset of its customers). Second, when the product itself \emph{depends} on causal estimation, scientific uncertainty is generally much higher, and so is the sophistication (and the associated investment) of the scientific work required.

\subsection{Example 3: GenAI Renovation Visualizations}
\label{sec:reno-viz}

Our third case applies the framework to a generative-AI product idea: rendering how a dated home would look after renovation.

Many on-market properties have not been recently renovated or refurbished. This can lower a prospective buyer's assessment of the home, especially if the buyer does not have much experience with modern renovations. The idea was to build an AI renovation-visualization tool, to help buyers see a dated property's potential. The main value assumption was that some listings fail to convert because buyers cannot see past the home's current condition. The tool would allow agents to show prospective buyers how the kitchen, say, would look after renovation. A worn kitchen with dated cabinetry and tired flooring could be compared with a newly renovated counterpart. The technical idea: use generative AI to create an image of how the kitchen would look after renovation.

\subsubsection{Value if Successful}

\paragraph{Strategic Alignment: Low to Medium.} The tool would help sell properties that might otherwise languish, supporting ``get more business,'' but would address a narrow slice of listings: those where cosmetic condition, not pricing or location, is the main barrier. Its direct monetary value to agents and the firm is minimal. If it worked, its stronger source of strategic value would lie in agents' interactions with prospective clients, helping differentiate them from other brokerages and thus win more business.

\paragraph{Direct Product Value: Low to Medium.} The upside here is small, perhaps \$50 million across 20,000 agents (an order of magnitude below Likely-to-Sell's roughly half a billion dollars), and even that figure is optimistic. One reasoning chain that gets us there picks up where the Time-on-Market tool left off: of the roughly 30\% of listings that do not convert, pricing explains perhaps three-quarters, leaving about 7.5\% of listings where property \emph{condition} is the main barrier. Optimistically, perhaps a third of those (roughly 2.5\% of all listings) involve homes attractive enough on other dimensions that a renovation visualization could tip a buyer's decision. 

With \compass{} agents averaging about four seller listings per year, that amounts to roughly 0.1 additional sold listings per agent per year. At approximately \$25,000 in commission revenue per listing, this yields roughly \$2,500 per agent annually, or the \$50 million total above. The figure is crude and sensitive to the assumed lift; treat it as a rough scale. Even \$50 million likely overstates the gain: agents already had a cheaper, accepted alternative (virtual staging/renovation\footnote{Available even in 2019 from third parties, though time-consuming and costly.}), so much of the value was already captured, leaving little \emph{incremental} gain, and no solid evidence the tool would lift agents' win rate with prospective clients.

\paragraph{Option Value: Low to Medium.} The generative rendering would likely lean on a third-party model, and thus would not create a proprietary data or model asset or platform component. There is some option value in the team improving its capability with generative AI for images.

\paragraph{Understanding Value: Low.} The project yields little insight into business dynamics or customer behavior. 

\paragraph{Overall Value Assessment: \textbf{Low to Medium}.}

\subsubsection{Likelihood of Success}

\paragraph{Product Certainty: Medium.} Because agents already had an accepted, lower-cost option in virtual staging, it was unclear how many would adopt a renovation-visualization tool on top of it. There was a further risk that the gap between the rendered visualization and the real property could \emph{increase} buyer disappointment on visiting, undermining trust in the agent and the listing.

\paragraph{Technical Certainty: Medium.} Generative models could produce convincing renovation renderings; however, they still sometimes made glaring errors. It was unclear whether those would scuttle the product. The core engineering carried little risk.

\paragraph{Data Certainty: High.} \compass{} had listing photos for all listed and to-be-listed properties. The generative models require minimal training data for the focal properties.

\paragraph{Scientific Certainty: Medium.} The image generation itself was feasible with contemporaneous models, but achieving consistently \emph{realistic and tasteful} results (appropriate to the property's price point, neighborhood, and architecture) required substantial prompt engineering and quality control. Unrealistic or generic-looking outputs could harm credibility.

\paragraph{Overall Likelihood Assessment: \textbf{Medium}.}

\subsubsection{Investment Required}

\paragraph{Data Science Team Effort: Medium.} When using a pre-trained model without fine-tuning, the data-science effort shifts to prompt engineering, increased complexity of system output evaluation, and API integration rather than ML model development.

\paragraph{Data Asset Investment: Low.} Existing listing photos suffice.

\paragraph{Integration and Productization Effort: High.} The tool needs careful UX design (including disclaimers), quality review workflows, agent training, and potentially legal review of generated images in the context of real-estate disclosures.

\paragraph{Overall Investment Assessment: \textbf{Medium}.} The single High on integration does not lift the overall rating, because the other two dimensions are Low (data assets) and Medium (team effort); the overall is a judgment call over the profile, not an average.

\subsubsection{The Upshot}

Despite the appeal of a GenAI-powered product, the renovation visualization tool ranks below the LTS product on nearly every dimension: lower value, lower likelihood, and comparable investment. Novelty does not translate into high \eroi{}. The framework asks decision-makers to judge GenAI opportunities by the same standard as traditional ones, not to favor a project simply for using the latest technology.

\compass{} did build a renovation visualization tool.\footnote{"Compass Lens," reported in the company's S1 filing.} However, as the risk from the GenAI errors was eventually deemed too large to deploy, the Compass Lens took a different approach, which in retrospect was a better product idea at the time. Specifically, given photos of a dated kitchen (for example), Compass Lens used AI to find very similar kitchens that had been renovated recently. It turned out that, in the huge database of homes on the market or recently sold, one could usually find a renovated kitchen remarkably similar in layout to the focal kitchen.

The product's reception was consistent with this modest assessment: it gained only limited traction, with many agents continuing to favor virtual staging, which met much of the same need at lower cost and risk. The episode illustrates that Direct Product Value must be judged against the best available alternative: the question is not whether a feature is valuable in a vacuum, but how much value it adds beyond what teams would do without it. 

\section{Conclusion: From Scores to Decisions}
\label{sec:conclusion}

The \eroi{} framework does not eliminate judgment; it structures it, making assumptions explicit and breaking the deadlock of needing to estimate ROI before knowing whether a project will work. Splitting one nebulous number into three lets a team see exactly where it agrees and where it disagrees, and prompts the conversation across business, product, engineering, and data science that a single score would bury. It complements recent guidance emphasizing strategy before technology \citep{hbr_ai_strategy_value} and capability-building through incremental wins \citep{smr_genai_smallt}: \eroi{} supplies the structure for choosing \emph{which} projects to fund. Whether the candidate is a traditional machine learning application, a generative-AI product, or a research investment, the same three questions apply: How valuable will it be if it succeeds? How likely is it to succeed? And what will it cost?

The template is short: rate the three pillars and their sub-dimensions on the coarse Low-to-Very-High scale using the anchors in Table~\ref{tab:rubric}, visualize the profile as in Figure~\ref{fig:eroi-visualization} so stakeholders can grasp it at a glance, and revisit the rankings as new information emerges. The step easiest to skip is one we have already argued for and the one that carries the most weight: build a portfolio rather than picking the single top-ranked project, funding a mix whose values, likelihoods, and costs balance one another instead of crowning one winner.

Our analysis has limits: as noted, although we used the framework at the time, the \compass{} case is retrospective and author-involved, so it illustrates how the framework organizes such choices rather than providing a test of its predictive accuracy. The business cases throughout are coarse, back-of-the-envelope estimates meant to expose the reasoning, not audited figures. A single company in a single industry cannot establish how far the framework generalizes. We have used the framework in other settings, and our students have even more broadly; even so, we expect its dimensions to be refined as others apply it in settings we have not seen.

A final caution concerns responsible deployment. The products in our case study are human-facing: Likely-to-Sell ranks an agent's personal contacts, and the renovation tool generates marketing imagery of real homes. Putting such systems into production raises questions the \eroi{} score does not settle. How is contact data obtained, consented to, and secured? Could targeting or generated imagery disadvantage particular neighborhoods or price tiers, a fair-housing concern in U.S. residential real estate? Are AI-generated renderings disclosed so that buyers are not misled? Many of these concerns are already inside the framework, in dimensions the earlier sections named: compliance review, fairness and privacy safeguards, and disclosure add to Integration and Productization Effort, and trust and safety risks lower Product Certainty, so a project that cannot be deployed responsibly is not high-\eroi{}. The concerns that do not fit a dimension we treat as a gating constraint on what reaches production, not an afterthought.

Few projects will match Likely-to-Sell's impact, but systematically choosing high-\eroi{} projects substantially improves the odds of AI investment success.

\section*{Author contributions}

Both authors contributed equally to this work. Under the CRediT taxonomy, both share the roles of Conceptualization, Writing -- original draft, and Writing -- review \& editing.

\section*{Statements and Declarations}

\paragraph{Ethical considerations.} This article does not report research involving human participants, human data, human tissue, or animals, so ethical approval was not required. The AI products discussed in the case study are nonetheless human-facing; we address their privacy, fair-housing, and disclosure implications in the Conclusion.

\paragraph{Consent to participate.} Not applicable; this article does not involve human participants.

\paragraph{Consent for publication.} Not applicable; this article does not contain data, images, or videos from any identifiable individual.

\paragraph{Declaration of conflicting interest.} The authors disclose the following potential conflicts of interest. Both authors were employees of \compass{} Inc.\ from 2019 to 2022, the period during which the AI initiatives described in this case study were undertaken. In addition, Panos Ipeirotis is listed as a member of the editorial board of \emph{Big Data}, though he is inactive in that role and has not served as a handling editor for the journal; he has no role in the editorial handling or peer review of this manuscript, which the journal manages independently in line with its policy for editorial-board submissions. The authors declare no other competing interests.

\paragraph{Funding statement.} The authors received no financial support for the research, authorship, and/or publication of this article.

\paragraph{Data availability.} This article presents a conceptual framework and a retrospective case study; it does not generate or analyze a research dataset. All quantitative figures are either drawn from the publicly available sources cited herein or are explicitly labeled back-of-the-envelope estimates, so no dataset is available to share.

\paragraph{Use of artificial intelligence.} The authors used large language models as assistive tools for copyediting and language refinement, and for editorial review and revision of the manuscript, under the authors' direction and final review. No generative AI was used to produce the research, its analysis, or its results, and no AI system is listed as an author.

\bibliographystyle{unsrtnat}
\bibliography{references}

@misc{NAR_profile,
  title={2023 Profile of Home Buyers and Sellers},
  author={Lautz, Jessica and others},
  year={2023},
  howpublished={National Association of {REALTORS}},
  url={https://www.nar.realtor/research-and-statistics/research-reports/highlights-from-the-profile-of-home-buyers-and-sellers}
}

@article{fernandez2022causal,
  title={Causal decision making and causal effect estimation are not the same… and why it matters},
  author={Fern{\'a}ndez-Lor{\'\i}a, Carlos and Provost, Foster},
  journal={INFORMS Journal on Data Science},
  volume={1},
  number={1},
  pages={4--16},
  year={2022},
  publisher={INFORMS},
  url={https://arxiv.org/abs/2104.04103}
}

@book{provost2013data,
  title={Data Science for Business: What You Need to Know about Data Mining and Data-Analytic Thinking},
  author={Provost, Foster and Fawcett, Tom},
  publisher={O'Reilly Media},
  year={2013},
  isbn={978-1449361327}
}

@article{mims2020ai,
  title={AI Isn't Magical and Won't Help You Reopen Your Business},
  author={Mims, Christopher},
  journal={The Wall Street Journal},
  year={2020},
  url={https://www.wsj.com/articles/ai-isnt-magical-and-wont-help-you-reopen-your-business-11590811201}
}

@misc{compass_lts_product,
  title={Likely to Sell Recommendations for Real Estate},
  author={{Compass True North}},
  year={2020},
  howpublished={Compass Engineering Blog},
  url={https://medium.com/compass-true-north/likely-to-sell-recommendations-for-real-estate-47e2f5c37f4}
}

@misc{compass_lts_ml,
  title={Machine Learning in Action for {Compass}'s Likely-to-Sell Recommendations},
  author={{Compass True North}},
  year={2020},
  howpublished={Compass Engineering Blog},
  url={https://medium.com/compass-true-north/machine-learning-in-action-for-compasss-likely-to-sell-recommendations-699a6dcd5076}
}

@misc{compass_earnings_2021q4,
  title={{Compass Inc.} {Q4} 2021 Earnings Call Transcript},
  author={{Compass Inc.}},
  year={2022},
  howpublished={Seeking Alpha},
  url={https://seekingalpha.com/article/4487666-compass-inc-comp-ceo-robert-reffkin-on-q4-2021-results-earnings-call-transcript},
  note={Open-access transcript: \url{https://www.fool.com/earnings/call-transcripts/2022/02/17/compass-inc-comp-q4-2021-earnings-call-transcript/}}
}

@misc{compass_fy2025,
  title={{Compass, Inc.} Reports Record Fourth Quarter and Full-Year 2025 Results},
  author={{Compass, Inc.}},
  year={2026},
  howpublished={Press release},
  url={https://www.prnewswire.com/news-releases/compass-inc-reports-record-fourth-quarter-and-full-year-2025-results-302698918.html}
}

@misc{mckinsey_genai,
  title={The Economic Potential of Generative {AI}: The Next Productivity Frontier},
  author={{McKinsey Global Institute}},
  year={2023},
  howpublished={McKinsey \& Company},
  url={https://www.mckinsey.com/capabilities/tech-and-ai/our-insights/the-economic-potential-of-generative-ai-the-next-productivity-frontier}
}

@techreport{mit_ai_pilots,
  title={The {GenAI} Divide: State of {AI} in Business 2025},
  author={Challapally, Aditya and Pease, Chris and Raskar, Ramesh and Chari, Pradyumna},
  institution={{MIT} Project {NANDA}},
  year={2025},
  month={July},
  url={https://nanda.media.mit.edu/ai_report_2025.pdf},
  urldate={2026-06-21}
}

@techreport{rand_ai_failure,
  title={The Root Causes of Failure for Artificial Intelligence Projects and How They Can Succeed: Avoiding the Anti-Patterns of {AI}},
  author={Ryseff, James and De Bruhl, Brandon F. and Newberry, Sydne J.},
  year={2024},
  institution={{RAND} Corporation},
  number={RR-A2680-1},
  url={https://www.rand.org/pubs/research_reports/RRA2680-1.html}
}

@misc{spglobal_ai_failure,
  title={Generative {AI} Shows Rapid Growth but Yields Mixed Results},
  author={{S\&P Global Market Intelligence}},
  year={2025},
  howpublished={Voice of the Enterprise: {AI} \& Machine Learning},
  url={https://www.spglobal.com/market-intelligence/en/news-insights/research/2025/10/generative-ai-shows-rapid-growth-but-yields-mixed-results}
}

@article{hbr_ai_strategy_value,
  title={Make Sure Your {AI} Strategy Actually Creates Value},
  author={Kim, W. Chan and Mauborgne, Ren\'{e}e and Ji, Mi},
  journal={Harvard Business Review},
  year={2025},
  month={September},
  url={https://hbr.org/2025/09/make-sure-your-ai-strategy-actually-creates-value}
}

@article{smr_llm_guide,
  title={A Practical Guide to Gaining Value from {LLM}s},
  author={Ramakrishnan, Rama},
  journal={MIT Sloan Management Review},
  volume={66},
  number={2},
  year={2024},
  url={https://sloanreview.mit.edu/article/a-practical-guide-to-gaining-value-from-llms/}
}

@article{smr_genai_smallt,
  title={Generate Value from {GenAI} with ``Small t'' Transformations},
  author={Webster, Melissa and Westerman, George},
  journal={MIT Sloan Management Review},
  year={2025},
  month={January},
  url={https://sloanreview.mit.edu/article/generate-value-from-gen-ai-with-small-t-transformations/}
}

@misc{smr_genai_exec_guide,
  title={Gaining Real Business Benefits from {GenAI}: An {MIT SMR} Executive Guide},
  author={{MIT Sloan Management Review}},
  year={2025},
  month={January},
  howpublished={MIT Sloan Management Review Executive Guide},
  url={https://sloanreview.mit.edu/article/gaining-real-business-benefits-from-genai-an-mit-smr-executive-guide/}
}

@article{siegel2024measuring,
  title={What Leaders Should Know About Measuring {AI} Project Value},
  author={Siegel, Eric},
  journal={MIT Sloan Management Review},
  year={2024},
  month={February},
  url={https://sloanreview.mit.edu/article/what-leaders-should-know-about-measuring-ai-project-value/}
}

@article{davenport2025enterprise_genai,
  title={How to Make Enterprise {Gen AI} Work},
  author={Valentine, Melissa and Politzer, Daniel J. and Davenport, Thomas H.},
  journal={Harvard Business Review},
  year={2025},
  month={September},
  url={https://hbr.org/2025/09/how-to-make-enterprise-gen-ai-work}
}

@misc{deloitte_genai_2025,
  title={Now Decides Next: The State of Generative {AI} in the Enterprise},
  author={{Deloitte}},
  year={2025},
  howpublished={Deloitte Global Survey},
  url={https://www.deloitte.com/az/en/issues/generative-ai/state-of-generative-ai-in-enterprise.html},
  note={Fourth-quarter wave, published January 2025}
}

@misc{bcg_value_ai2024,
  title={Where's the Value in {AI}?},
  author={{Boston Consulting Group}},
  year={2024},
  howpublished={BCG},
  url={https://www.bcg.com/publications/2024/wheres-value-in-ai}
}

@inproceedings{amdahl1967,
  title={Validity of the Single Processor Approach to Achieving Large Scale Computing Capabilities},
  author={Amdahl, Gene M.},
  booktitle={Proceedings of the April 18--20, 1967, Spring Joint Computer Conference (AFIPS)},
  pages={483--485},
  year={1967},
  doi={10.1145/1465482.1465560}
}

@misc{rice_mcbride2018,
  title={{RICE}: Simple Prioritization for Product Managers},
  author={McBride, Sean},
  year={2018},
  howpublished={Intercom Blog},
  url={https://www.intercom.com/blog/rice-simple-prioritization-for-product-managers/}
}

@book{ice_ellis2017,
  title={Hacking Growth: How Today's Fastest-Growing Companies Drive Breakout Success},
  author={Ellis, Sean and Brown, Morgan},
  publisher={Currency},
  address={New York},
  year={2017},
  isbn={978-0451497215}
}

@book{realoptions_dixit1994,
  title={Investment under Uncertainty},
  author={Dixit, Avinash K. and Pindyck, Robert S.},
  publisher={Princeton University Press},
  address={Princeton, NJ},
  year={1994},
  isbn={978-0691034102}
}

@book{realoptions_trigeorgis1996,
  title={Real Options: Managerial Flexibility and Strategy in Resource Allocation},
  author={Trigeorgis, Lenos},
  publisher={MIT Press},
  address={Cambridge, MA},
  year={1996},
  isbn={978-0262201025}
}

@article{stagegate_cooper1990,
  title={Stage-Gate Systems: A New Tool for Managing New Products},
  author={Cooper, Robert G.},
  journal={Business Horizons},
  volume={33},
  number={3},
  pages={44--54},
  year={1990},
  publisher={Elsevier},
  doi={10.1016/0007-6813(90)90040-I}
}

@book{cooper_portfolio2001,
  title={Portfolio Management for New Products},
  author={Cooper, Robert G. and Edgett, Scott J. and Kleinschmidt, Elko J.},
  edition={2nd},
  publisher={Perseus Publishing},
  address={Cambridge, MA},
  year={2001},
  isbn={978-0738205144}
}

@article{howard1988,
  title={Decision Analysis: Practice and Promise},
  author={Howard, Ronald A.},
  journal={Management Science},
  volume={34},
  number={6},
  pages={679--695},
  year={1988},
  publisher={INFORMS},
  doi={10.1287/mnsc.34.6.679}
}

@article{dimasi2003,
  title={The Price of Innovation: New Estimates of Drug Development Costs},
  author={DiMasi, Joseph A. and Hansen, Ronald W. and Grabowski, Henry G.},
  journal={Journal of Health Economics},
  volume={22},
  number={2},
  pages={151--185},
  year={2003},
  publisher={Elsevier},
  doi={10.1016/S0167-6296(02)00126-1}
}

@article{dimasi2001,
  title={Risks in New Drug Development: Approval Success Rates for Investigational Drugs},
  author={DiMasi, Joseph A.},
  journal={Clinical Pharmacology \& Therapeutics},
  volume={69},
  number={5},
  pages={297--307},
  year={2001},
  publisher={Wiley},
  doi={10.1067/mcp.2001.115446}
}

@article{holland1986,
  title={Statistics and Causal Inference},
  author={Holland, Paul W.},
  journal={Journal of the American Statistical Association},
  volume={81},
  number={396},
  pages={945--960},
  year={1986},
  publisher={Taylor \& Francis},
  doi={10.1080/01621459.1986.10478354}
}

@book{imbens_rubin2015,
  title={Causal Inference for Statistics, Social, and Biomedical Sciences: An Introduction},
  author={Imbens, Guido W. and Rubin, Donald B.},
  publisher={Cambridge University Press},
  address={New York},
  year={2015},
  isbn={978-0521885881},
  doi={10.1017/CBO9781139025751}
}

@book{angrist_pischke2009,
  title={Mostly Harmless Econometrics: An Empiricist's Companion},
  author={Angrist, Joshua D. and Pischke, J{\"o}rn-Steffen},
  publisher={Princeton University Press},
  address={Princeton, NJ},
  year={2009},
  isbn={978-0691120355}
}

\end{document}